\documentclass[%
 reprint,
 superscriptaddress,
 amsmath,amssymb,
 aps,
 prd,
]{revtex4-1}

\usepackage{dcolumn}% Align table columns on decimal point
\usepackage{epsfig}
\usepackage{subfigure}
\usepackage{bm}% bold math
\usepackage{breakurl}
\usepackage{enumitem}
\usepackage{mdframed}

\newcommand\fig[1]     {Fig.\,{\ref{#1}}}
\usepackage{color}

\newcommand{\beq}{\begin{equation}}
\newcommand{\eeq}{\end{equation}}
\newcommand{\beqa}{\begin{eqnarray}}
\newcommand{\eeqa}{\end{eqnarray}}
\newcommand\eq[1]     {(\ref{#1})}

\begin{document}

\title{Thermal and Quantum Phase Transitions of the $\phi^4$ Model}

\author{Istv\'an G\'abor M\'ari\'an}
\affiliation{HUN-REN Atomki, P.O.Box 51, H-4001 Debrecen, Hungary} 
\affiliation{University of Debrecen, Institute of Physics, P.O.Box 105, H-4010 Debrecen, Hungary}

\author{Andrea Trombettoni}
\affiliation{Department of Physics, University of Trieste, Strada Costiera 11, I-34151 Trieste, Italy}
\affiliation{CNR-IOM DEMOCRITOS Simulation Center, Via Bonomea 265, I-34136 Trieste, Italy}

\author{Istv\'an N\'andori}
\affiliation{University of Miskolc, Institute of Physics and Electrical Engineering, H-3515, Miskolc, Hungary}
\affiliation{HUN-REN Atomki, P.O.Box 51, H-4001 Debrecen, Hungary} 

\date{\today}

\begin{abstract}
In this paper we discuss and revisit the finite temperature extension of the renormalization group 
(RG) treatment of $T=0$ field theories, focusing as a case study on the $\phi^4$ model. We first discuss the 
extension of RG equations of the very same model from $T=0$ to finite $T$ in the usual way by resorting to 
sums on the Matsubara frequencies and fixing the physical temperature parameter $T$. We show that this 
approach, although useful for a variety of applications, may lead to the disappearance of the critical points 
as extracted from the RG flow. Since the identification of fixed points is key in the study of classical and 
quantum phase transitions, we
propose a modification of the usual finite-temperature RG approach by relating the temperature parameter 
to the running RG scale, $T \equiv k_T = \tau k$ where $k_T$ is the running cutoff for thermal, and $k$ is 
for the quantum fluctuations. Once introduced this dimensionless temperature $\tau$, we investigate the 
consequences on the thermal RG approach for the $\phi^4$ model and construct its phase diagram.
Finally, we formulate requirements for the phase diagram of the $\phi^4$ theory based on known properties of 
the quantum and classical phase diagrams of the Ising model. 
\end{abstract}

\maketitle

%===============
\section{Introduction} 
%===============
A quantum phase transition (QPT) occurs at the quantum critical point at zero temperature 
where thermal fluctuations are absent. The QPT is understood as a transition between 
different quantum phases and driven by quantum fluctuations. Contrary to a classical phase 
transition (CPT) which is a thermal phase transition, the QPT is accessed at 
absolute zero temperature by varying a physical parameter such as the magnetic field or the 
pressure. In this work, we refer to this as the 'quantum' parameter. 
As a major example, the critical behaviour of the CPT of the Ising model is in the 
same universality class of the QPT of the $\phi^4$ Quantum Field Theory (QFT), this providing 
a paradigmatic instance of the relation between a CPT and its quantum counterpart 
QPT. It is well known that a classical system in $d$ dimension is --  for short-range interactions -- 
in the same universality class of the same model in $d-1$ spatial dimension at $T=0$.

At a finite but suitably low temperature, 
classical and quantum fluctuations compete with each other.
In the quantum critical region near the quantum critical point, quantum
fluctuations dominate the system behaviour and
one can study properties of the QPT \cite{sachdev}.
Thus, a so called QPT-CPT phase diagram can be 
drawn where the ordered and the disordered phases are separated by each other by a 
critical line which connects the critical temperature and the quantum critical point. 
Around the classical phase transition, in the classical critical region,
the system is governed by
thermal fluctuations and this region becomes narrower with decreasing temperatures 
and converges towards the quantum critical point monotonically. 

The Ising model in $d=2$ or $d=3$ dimensions represents an excellent playground to 
test and to understand the interplay between quantum and classical phase transitions. 
The quantum Ising model \cite{sachdev,tranverse_ising_1,tranverse_ising_2} can be 
realized by adding a transverse, external magnetic field to the usual Hamiltonian
of the classical Ising model. 
This is the so called transverse field Ising model where the QPT is realized by
varying 
the magnitude of the transverse external magnetic field \cite{tranverse_ising_1,tranverse_ising_2}. 
The QPT-CPT phase diagram of the Ising model is well-known and its interpretation in 
terms of effective field theories also deeply investigated \cite{sachdev,shankar}.
The classical Ising
model with vanishing external magnetic field undergoes a second order classical (thermal)
phase transition at a finite critical temperature $T_c$. The quantum Ising model at zero 
temperature has a QPT at a critical magnetic field $h_c$. 

It is also well-known that the classical Ising model (without any transverse magnetic field) 
can be mapped onto the $\phi^4$ Quantum Field Theory (QFT) which has a QPT where 
its quantum critical behaviour is found to be identical to the classical critical behaviour of 
the Ising model. As a consequence, the two models belong to the same universality class 
and their critical exponents are identical. A natural question is what is the 
QPT-CPT diagram of the $\phi^4$ QFT and how it is related to that of the Ising model,
an issue that has been investigated using a variety of tools.
In this article we aim to clarify this issue using the Functional Renormalization Group approach
(FRG) \cite{eea_rg}, where we need to compare results obtained at zero temperature to
those obtained when temperature is introduced. We will argue that a {\it dimensionless} 
temperature has to be introduced, as we discuss in the next section.

%===============
\section{Main idea}
%===============

In general, in order to map out the QPT-CPT diagram of the $\phi^4$ field theory one has to 
perform its thermal RG analysis where both quantum and thermal fluctuations are
taken into account. This can be done by e.g., using the finite temperature extension 
of the zero-temperature FRG method. The standard formulation of the FRG method \cite{eea_rg} 
is written in Euclidean spacetime in the framework of the zero-temperature QFT. Its 
generalization to finite temperature requires the inclusion of the inverse temperature 
parameter $\beta = 1/k_B T$ as the upper bound for the imaginary time integral
\cite{rg_dimful_T_1,rg_dimful_T_2,finite_T_V_1,thermal_rg_phi4,thermal_rg_phi4_pressure,
rg_dimful_T_5,rg_dimful_T_6,rg_dimful_T_8,finite_T_V_3,rg_dimful_T_10,finite_T_V_5,
thermal_rg_phi4_volume,rg_dimless_T_1,rg_dimless_T_2,rg_dimless_T_3,rg_dimless_T_4,
rg_dimless_T_5,rg_dimless_T_6,rg_dimless_T_7,rg_dimless_T_8,rg_dimless_T_9,
rg_dimless_T_10,rg_dimless_T_11}.

Once $\beta$ has been introduced, one has to relate the temperature parameter
to a momentum scale by using units in which $c = \hbar = k_B = 1$.
In the usual perturbative approach the standard choice is $\mu = 2\pi T$ where $\mu$ 
is the perturbative RG scale which is related to the non-perturbative one: $\mu \sim k$, 
where $k$ is the momentum scale. However, 
it is known that the $k$-dependence of the effective action of the FRG equation is 
only introduced artificially to implement the Wilsonian integration \cite{wilson} of fluctuating 
modes and the quantised theory is obtained in the physical limit $k \to 0$.
Thus, one has to take the limit $k\to 0$. For this reason, in Refs.~\cite{rg_dimful_T_1,
rg_dimful_T_2,finite_T_V_1,thermal_rg_phi4,thermal_rg_phi4_pressure,rg_dimful_T_5,
rg_dimful_T_6,rg_dimful_T_8,finite_T_V_3,rg_dimful_T_10,finite_T_V_5,thermal_rg_phi4_volume}
the running couplings are defined at an arbitrary but fixed intermediate momentum 
scale $k_\star = 2\pi T = 2\pi \tau \Lambda$ where $\Lambda$ is the UV initial value 
of the running momentum cutoff $k$. So, the temperature parameter is linked to
the UV cutoff, i.e.,~$T = \tau \Lambda$. Once the couplings are defined, the 
zero-temperature FRG equation is integrated from $k_\star$ up to $\Lambda$ and 
then starting from these bare parameters, the FRG equation is solved down from 
$k=\Lambda$ to $k=0$ but this time with the temperature $T$ turned on and the 
physical quantities are obtained at $k=0$. This investigation is then performed
with the usual tools of QFT.

However, every approach has advantages and disadvantages.
In the fixed $T$ approach the dimensionful temperature is well-defined, the IR limit $k \to 0$ 
can be taken safely but the RG flow equations have no fixed point solutions, since
the explicit $k$-dependence cannot be removed from the dimensionless RG flow 
equations. For example, the dimensionless thermal FRG equation (3.9) of 
\cite{thermal_rg_phi4} which is given in the so called local potential approximation, 
has an explicit $k$-dependence and it blows up in the IR limit, i.e., for $k \to 0$ 
where the attractive fixed points are always given, so it makes no room 
for non-trivial fixed points such as the Wilson-Fisher (WF) one. 
As an example, we mention that in Refs.~\cite{rancon_1,rancon_2,rancon_3} the quantum 
$O(N)$ and the Bose-Hubbard model were studied at finite temperatures and it was shown
that in order to properly identify the "quantum WF fixed point" (i.e. at zero-temperature) and 
the "classical WF fixed point" (at finite temperatures, where quantum fluctuations are irrelevant), 
one needs to use different dimensionless coupling constants, see more in the following.

In this work we %make an attempt to
deal with the above problems by suggesting the 
identification $$T \equiv k_T = \tau k$$ where $k_T$ is the running cutoff for thermal, 
and $k$ is for the quantum fluctuations. This choice can be supported by the following 
argument. The Wilsonian approach requires the rescaling of the upper bound of the 
imaginary time integral in every blocking step, so, it has to be linked to the running 
momentum cutoff $k$. However, in the FRG method one has to take the limit $k \to 0$, 
therefore, the upper bound cannot be considered as the (inverse) temperature but can 
be used as a running momentum cutoff $T\equiv k_T$. Thus, the temperature is related 
to the dimensionless quantity $\tau$ which is kept constant over the RG flow. A dimensionless 
(reduced) temperature has been used in the framework of FRG \cite{rg_dimless_T_1,
rg_dimless_T_2,rg_dimless_T_3,rg_dimless_T_4,rg_dimless_T_5,rg_dimless_T_6,
rg_dimless_T_7,rg_dimless_T_8,rg_dimless_T_9,rg_dimless_T_10,rg_dimless_T_11}, 
however, in these works the typical choice is $\tau \equiv \tau_k = T/k$ which means that
the dimensionful parameter $T$ is kept constant over the RG flow, so, $\tau_k$ has a 
trivial (non-vanishing) dependence on the RG scale $k$, see \cite{details}
for a few examples. One may summarize that the dimensionless temperature is
introduced as a technical tool in the literature 
\cite{rg_dimless_T_1,rg_dimless_T_2,rg_dimless_T_3,rg_dimless_T_4,rg_dimless_T_5,rg_dimless_T_6,rg_dimless_T_7,rg_dimless_T_8,rg_dimless_T_9,rg_dimless_T_10,rg_dimless_T_11}, 
yet a complete study of consequences and the determination of physical properties from the
{\em constant}  dimensionless temperature is lacking, to the best of our knowledge,
making it more difficult to understand and to identify thermal and quantum phase transitions and their crossovers, 
as we are going to argue in the following. 

It is important to note that our proposal for fixing $\tau = T/k$ and taking the simultaneous 
$T, k \to 0$ limit addresses a complementary regime compared to the widely used fixed $T$ 
approach. In a certain level of approximation the two methods can give qualitatively same 
results, so one cannot come to the conclusion that previous studies were incorrect. However, 
it is also important to note that in our fixed $\tau$ approach the RG flow equations have real 
(and not pseudo) fixed points which have fundamental importance since the determination of 
critical behaviour and various phases are strongly related to fixed point solutions of RG flow 
equations. 

Thermal field theory is a very well-developed framework for finite temperature 
applications \cite{kapusta} and it is employed in many-body physics \cite{negele,kapusta} 
which relates a quantum system in $d-1$ (spatial) dimensions at $T=0$ and the 
corresponding classic system in $d$ dimensions at finite temperature \cite{sachdev}. 
It generates the discretisation of the imaginary time-integral in momentum space, i.e., 
summation on Matsubara frequencies \cite{negele,kapusta}. At non-zero temperatures, 
classical fluctuations with an energy scale of $ k_B T$ compete with the quantum 
fluctuations of energy scale $\hbar \omega$ where $\omega$ is the frequency of quantum 
oscillations. By using natural units, one finds that $\omega$ must be compared to $T$ and
in the RG method one has to relate the frequency $\omega$ to the running cutoff  $k$ 
which is used to integrate quantum fluctuations systematically, so one finds $\omega =k$.
Thus, one can introduce a dimensionless quantity, $\tau = T/k = T/\omega$ which is 
used to compare the strength of quantum and thermal fluctuations. If $\tau \lesssim 1$ then 
quantum fluctuations dominate, if $\tau \gtrsim 1$ then thermal fluctuations dominate.
Notice that in the Wilsonian approach fluctuations are taken into account by the successive 
elimination (integration) of degrees of freedom above these running cutoffs which are chosen 
to be different in order to make difference between thermal and quantum fluctuations. In other 
words, for $k_T \sim k$ (i.e.,~for $\tau  \sim 1$) one cannot distinguish between thermal and 
quantum fluctuations. 

We aim to explore the possibility of a thermal RG approach in the framework of the FRG 
method using the identification $T = \tau k$ where the temperature is related to the 
dimensionless quantity $\tau$ which is kept constant over the RG flow. Once the 
dimensionless temperature $\tau$ has been introduced, we aim at determining
the consequences of this introduction in the FRG formalism and how the points 
previously discussed can dealt with. We apply this new thermal RG approach for the 
$\phi^4$ model. We are motivated in this study by the results presented
in Ref.~\cite{thermal_rg_cosmo} where the thermal RG method for the 4-dimensional 
$\phi^4$ theory has been studied and its consequences on key issues of Inflationary 
Cosmology investigated.

Our goal here is to map out the QPT-CPT phase diagram of the $\phi^4$ theory
and compare it to the QPT-CPT diagram of the Ising model. It represents graphically the interplay 
between the classical phase transition (CPT) and the quantum phase transition (QPT) of the $\phi^4$ 
scalar field theory which has also been investigated in connection to the Naturalness/Hierarchy 
problem \cite{BrBrCoDa,BrBrCo,qpt_higgs1,qpt_higgs2}. For example, in \cite{qpt_higgs1,qpt_higgs2} 
it was shown that the hierarchy problem as well as the metastability of the electroweak vacuum can 
be understood as the Higgs potential being near-critical, i.e., close to a QPT. We 
discuss known properties of the thermal and quantum phase transitions of the 
Ising model and based on these, we formulate requirements for the QPT-CPT 
phase diagram of the $\phi^4$ theory. We construct the QPT-CPT phase diagram 
of the $\phi^4$ theory (focusing to dimensions below the upper critical dimension)
by the thermal RG approach suggested 
in this work and compare it to the formulated requirements in order to check the 
viability of the new method.

%===============
\section{Modified thermal RG equation} 
%===============
The FRG equation \cite{eea_rg} at zero-temperature is formulated in Euclidean spacetime
and reads as
\begin{eqnarray}
\label{erg}
k \partial_k \Gamma_k[\phi] = \frac{1}{2} 
\int \frac{d^d p}{(2\pi)^d} \,
\frac{k \, \partial_k R_k(p)}{R_k(p) + \Gamma^{(2)}_k[\phi]}, 
\end{eqnarray}
where $\Gamma_k[\phi]$ is the running effective action with its Hessian $\Gamma^{(2)}_k[\phi]$
and $k$ is the running momentum cutoff (i.e., the RG scale) and $R_k(p)$ is an appropriately 
chosen regulator function.  In order to handle the FRG equation one has to use approximations. 
In the so called gradient expansion, the action is expanded in terms of the derivatives of the field. 
The lowest order of the gradient expansion is the so called, local potential approximation (LPA). 
In LPA when the couplings of the scaling potential $V_k(\phi)$ carry RG scaling only, the 
Wetterich FRG equation \eq{erg} is written as
\begin{eqnarray}
\label{lpa_erg}
k \partial_k {V_k(\phi)} =  \frac{1}{2} \int_{-\infty}^{\infty}  \frac{d^d p}{(2\pi)^d} \, 
\frac{k\partial_k R_k}{p^2 + R_k + V''_k(\phi) } \,,
\end{eqnarray}
with $V''_k = \partial^2_\phi V_k$. 

The extension of the zero-temperature FRG method for finite temperatures requires the 
modification of the imaginary time $\tilde t$ integral $\int d^dx \to \int_0^\beta d\tilde t \int d^{d-1}x$ 
where $\beta = 1/T$ and $T$ is the temperature parameter. As a next step, the momentum integral 
with respect to the imaginary time is modified $\int d^dp \to T \sum_{\omega_n} \int d^{d-1}p$ 
by using Matsubara frequencies where the summation is performed over these discrete
frequencies; for bosonic degrees of freedom $\omega_n = 2 \pi n T$.

%===============
\subsection{Thermal RG with the Litim regulator} 
%===============
In order to proceed further one has to assume that the regulator function is independent 
of the Matsubara frequency, but otherwise the same as for the zero-temperature case. This 
is the case of the "frequency independent regulator". In addition, it is also important to note 
that once approximations are used, the concrete form of the FRG equation start to depend 
on the particular choice of the regulator function. There are two types of regulator functions
where the momentum integrals in \eq{lpa_erg} and its finite-temperature counterpart 
can be performed analytically in LPA. The first one is the Litim regulator \cite{Litim2000}, 
which has the following form at zero-temperature,
\beq
\label{litim_regulator}
R_k(p) = (k^2 - p^2) \theta(k^2 - p^2) \,,
\eeq
where $\theta(y)$ is the Heaviside step function. The frequency-independent form of the 
Litim regulator results in the following thermal RG equation in LPA  \cite{thermal_rg_phi4},
\beq
\label{thermal_opt}
k\partial_k V_k(\phi) = \frac{2 \alpha_{d-1}}{d-1} \, k^{d-1} 
T \sum_{n=-\infty}^{\infty} \frac{k^2}{k^2+ \omega_n^2 +\partial^2_{\phi} V_k(\phi)} \,,
\eeq
with $\omega = 2\pi n T$ and $\alpha_d = \Omega_d/(2(2\pi)^d)$, where
$\Omega_d = 2 \pi^{d/2}/\Gamma(d/2)$ is the $d$-dimensional solid angle. 
The summation can be performed \cite{thermal_rg_phi4} which results in an 
RG equation (identical to Eq.(3.9) of \cite{thermal_rg_phi4} for $N=1$) 
\begin{eqnarray}
k\partial_k V_k(\phi) = \frac{2 \alpha_{d-1}}{d-1} \, k^{d+1}
\frac{2 n(\omega_k) +1}{2 \omega_k} \,,
\nonumber
\end{eqnarray}
where $n(\omega_k) = (\exp{(\omega_k/T)}-1)^{-1}$ is the bosonic distribution function 
with $\omega_k = \sqrt{k^2+\partial^2_{\phi} V_k(\phi)}$ which appears in the so 
called "thermal contribution" which vanishes in the limit $T\to 0$ and the 
remaining term is the so called "vacuum contribution". This equation can be 
rewritten as,
\begin{eqnarray}
k\partial_k V_k(\phi) = \frac{\frac{2 \alpha_{d-1}}{d-1} \, k^{d+1}}{2\sqrt{k^2+\partial^2_{\phi} V_k(\phi)}}
\frac{\exp{\left(\frac{\sqrt{k^2+\partial^2_{\phi} V_k(\phi)}}{T}\right)}+1}{\exp{\left(\frac{\sqrt{k^2+\partial^2_{\phi} V_k(\phi)}}{T}\right)} -1}
\nonumber
\end{eqnarray}
and can be further simplified by using the  identity $\coth(x/2) = \frac{\exp(x)+1}{\exp(x)-1}$ 
which results in,
\begin{eqnarray}
\label{thermal_opt_rg}
k\partial_k V_k(\phi)  = \frac{2 \alpha_{d-1}}{d-1} \, k^{d+1} 
\frac{\coth{\left(\frac{\sqrt{k^2+\partial^2_{\phi} V_k(\phi)}}{2T}\right)}}{2\sqrt{k^2+\partial^2_{\phi} V_k(\phi)}} \,.
\end{eqnarray}
Let us note, the above thermal RG equation is well-known. For example, in Ref.~\cite{thermal_rg_phi4}
the authors performed a detailed analysis of the $O(N)$ scalar field theory by the thermal RG equation 
\eq{thermal_opt_rg} which was also discussed in \cite{thermal_rg_phi4_pressure} with the inclusion 
of pressure and in \cite{thermal_rg_phi4_volume} by taking into account volume fluctuations too
where the temperature is linked to the UV cutoff $\Lambda$, i.e., $T = \tau \Lambda$.

\begin{figure}
\centering
\includegraphics[width=\linewidth]{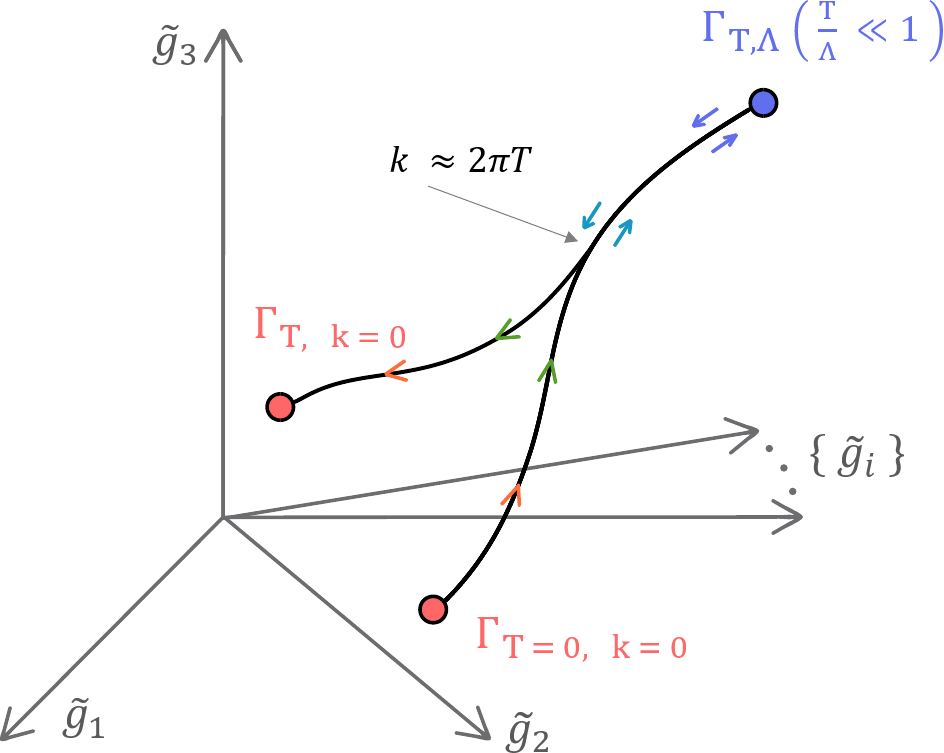}
\caption{
Schematic thermal RG flow with $T = \tau \Lambda$. 
The figure is similar to those found in \cite{rg_dimful_T_5,rg_dimful_T_6}.
}
\label{fig1} 
\end{figure}

The schematic thermal RG flow with $T = \tau \Lambda$ was shown in Fig.~6 of \cite{rg_dimful_T_5} 
and Fig.~4 of \cite{rg_dimful_T_6} and given here on \fig{fig1}.
%
% Fig 1
%
It is clear from \fig{fig1} that the zero-temperature FRG equation is integrated from 
$k = 0$ up to $\Lambda$ while the couplings are defined at an arbitrary but fixed 
intermediate momentum scale $k_\star = 2\pi T = 2\pi \tau \Lambda$. Then starting 
from these bare parameters, the thermal ($T\neq 0$) FRG equation is solved from 
$k=\Lambda$ to $k=0$. It is also clear, that RG flows at zero and at finite temperatures 
do not differ from each other for $k > k_\star$ which is evident from \eq{thermal_opt_rg} 
since in the limit $x \to \infty$ one finds $\coth(x) \to 1$. However, for $k < k_\star$ they 
start to deviate from each other, so, the physical quantities can be obtained at $k=0$ 
of the thermal RG flow. Nevertheless, this procedure has a drawback, the explicit 
$k$-dependence cannot be removed from the dimensionless RG flow equations.

In this work we use the identification $T \equiv k_T = \tau k$ where "$\tau$" plays the 
role of the (dimensionless) temperature which distinguishes between thermal 
quantum fluctuations. We suggest a modified thermal RG equation,
\beq
\label{thermal_opt_rg_k}
k\partial_k V_k(\phi) = \frac{2 \alpha_{d-1}}{d-1} \, k^{d+1} 
\frac{\coth{\left(\frac{\sqrt{k^2+\partial^2_{\phi} V_k(\phi)}}{2\tau k}\right)}}{2\sqrt{k^2+\partial^2_{\phi} V_k(\phi)}} \,,
\eeq
which was given in Ref.~\cite{thermal_rg_cosmo}, too. The advantage of this modified
thermal RG equation is the absence of $k$-dependence in the dimensionless RG flow 
equations.

%===============
\subsection{Thermal RG with the sharp-cutoff regulator} 
%===============
As we argued, in order to perform the momentum integrals in \eq{lpa_erg} and its 
finite-temperature counterpart analytically one has to choose a special type of regulator.
In the previous subsection we discussed the case of the Litim regulator \cite{Litim2000}. 
In this subsection we choose the sharp-cutoff which has the following form at zero
temperature,
\beq
\label{sharp_regulator}
R_k(p) = p^2 \left(\frac{1}{\theta(p^2-k^2)} -1 \right)
\eeq 
and it is considered as the simplest regulator function. The frequency-independent 
form of the sharp-cutoff regulator results in the following thermal RG equation in LPA,
\begin{eqnarray}
\label{thermal_sharp}
k\partial_k V_k(\phi)  &=& - \alpha_{d-1} \, k^{d-1} T \\
&&\sum_{n=-\infty}^{\infty} 
\ln \left( \frac{k^2+ \omega_n^2 +\partial^2_{\phi} V_k(\phi)}{k^2+ \omega_n^2 } \right) \,,
\nonumber
\end{eqnarray}
where the denominator inside the logarithm can be chosen with some freedom but it 
must have the dimension of momentum square and it has to be field-independent. 
For example, it could be $k^2$ or $T^2$. Here we take the choice when it becomes
identical to the nominator in the absence of fields. In this case, the summation can be 
performed which results in a thermal RG equation 
\begin{eqnarray}
\label{thermal_sharp_rg}
k\partial_k V_k(\phi) &=& - \alpha_{d-1} \, k^{d-1} 2 T  \\
&&\ln \left[ \rm{csch} \left( \frac{k}{2 T} \right) \sinh \left( \frac{ \sqrt {k^2 + \partial^2_{\phi} V_k(\phi)}  }{2 T} \right) \right] \,,
\nonumber
\end{eqnarray}
and by using the identification $T \equiv k_T = \tau k$ the following modified thermal 
RG equation can be obtained,
\begin{eqnarray}
\label{thermal_sharp_rg_k}
k\partial_k V_k(\phi) &=& - \alpha_{d-1} \, k^{d} 2 \tau   \\
&&\ln \left[ \rm{csch} \left( \frac{1}{2 \tau} \right) \sinh \left( \frac{ \sqrt {k^2 + \partial^2_{\phi} V_k(\phi)}  }{2 \tau k} \right) \right] \,.
\nonumber
\end{eqnarray}
%

%===============
\section{The zero-temperature limit} 
%===============
Before we start to apply the modified thermal RG equations \eq{thermal_opt_rg_k} 
and \eq{thermal_sharp_rg_k} let us discuss in this section an important issue, the 
zero-temperature limit. As we argued, the finite-temperature approach 
requires the modification of the imaginary time integral by introducing an upper bound which 
is the inverse temperature parameter, $\beta = 1/T$. This results in the replacement of the 
continuous (imaginary time) integral by a sum over discrete Matsubara frequencies in the 
momentum space in RG equations \eq{erg} and \eq{lpa_erg}. Here, the zero-temperature 
limit,  i.e., $\beta \to \infty$ must recover the RG equations \eq{erg} and \eq{lpa_erg}. 

However, if one would like to perform the momentum integrals (and the summation) the use 
of the regulator function is unavoidable. The finite-temperature formalism requires a frequency 
independent regulator which does not regulate the Matsubara sum. This is not true for the 
zero-temperature case where momentum integrals (including the imaginary time direction) 
are performed by the regulator function (i.e., with a frequency-dependent regulator). Thus, 
the zero-temperature limit of RG equations \eq{thermal_opt_rg} and \eq{thermal_sharp_rg} 
may differ from their counterparts at zero temperature.

Indeed, by choosing the frequency independent Litim regulator and taking the limit of 
zero-temperature, the thermal RG equation \eq{thermal_opt_rg} reduces to,
\beq
\label{opt_T}
{\rm T \to 0:} \hskip 0.5cm
k\partial_k V_k(\phi) = \frac{2 \alpha_{d-1}}{d-1} \frac{k^{d+1} }{2\sqrt{k^2+\partial^2_{\phi} V_k(\phi)}} \,,
\eeq
which is not identical to its zero-temperature counterpart
\beq
\label{opt}
{\rm T=0:} \hskip 0.5cm
k\partial_k V_k(\phi) = \frac{2 \alpha_d}{d} \, \frac{k^{d+2}}{k^2+\partial^2_{\phi} V_k(\phi)} \,,
\eeq
where the frequency-dependent Litim regulator is used to perform the momentum integrals
of \eq{lpa_erg}. The situation is similar for the sharp cutoff case, where the zero-temperature
limit of the thermal RG equation \eq{thermal_sharp_rg} is not identical to its zero-temperature 
counterpart, i.e., the Wegner-Houghton equation \cite{wh_rg} at LPA.

The essence of this problem is whether the frequency dependence is taken separately or not 
which is leading to cylindrically (frequency-dependent regulator) or spherically (frequency 
independent regulator) symmetric geometry \cite{cylindrical_symm_1,cylindrical_symm_2}. 
The cylinder can have finite or infinite size. An infinite size cylinder with the Litim regulator
results in an FRG equation (see Eq.~(7) of \cite{cylindrical_symm_2}) identical to \eq{opt_T}.

It is clear that equations \eq{opt_T} and \eq{opt} have different forms, however, they have 
the same singularity structure ($k^2+\partial^2_{\phi} V_k = 0$) which results in a convex 
dimensionful potential in the IR limit ($k\to 0$) in both cases. The position and the critical 
behaviour of the WF fixed point obtained by equations \eq{opt_T} and \eq{opt}
may differ from each other, i.e., the zero-temperature limit requires attention. Actually, fixed
points are not connected to physical observables, but the critical exponents which are used 
to characterise the critical behaviour around the WF fixed point can be measured directly,
thus, they can be used to study the zero-temperature limit problem.

One of our goals in this work is to use measurable quantities, i.e., critical exponents to 
study the zero-temperature limit of the thermal RG approach by using the modified RG 
equations \eq{thermal_opt_rg_k} and \eq{thermal_sharp_rg_k}

%===============
\section{Thermal RG study of the $\phi^4$ model in lower dimensions} 
\begin{figure}
\centering
\includegraphics[width=\linewidth]{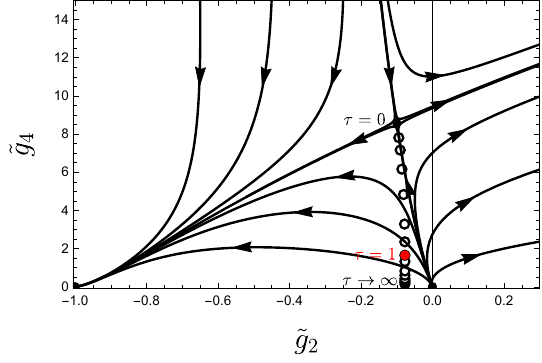}
\caption{
Thermal RG flow diagram of the $\phi^4$  model in $d=3$ dimensions based on 
flow equations \eq{g2_flow} and \eq{g4_flow}. Circles show how the position 
of the WF fixed point changes with $\tau$.
}
\label{fig2} 
\end{figure}

\begin{figure}
\centering
\includegraphics[width=\linewidth]{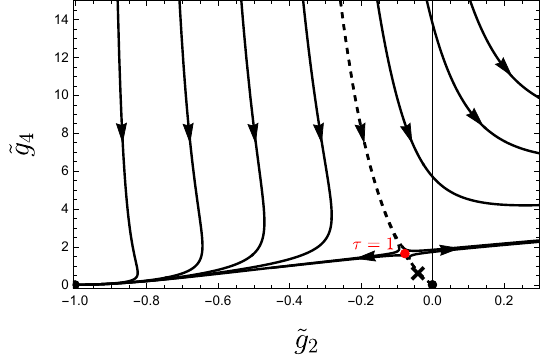}
\caption{
Thermal RG flow diagram of the $\phi^4$  model in $d=3$ dimensions with 
$\tau = 1$ based on the RG flow equations \eq{g2_flow} and \eq{g4_flow}. 
Black cross denotes an initial condition which lies on the separatrix. For 
$\tau < 1$ the RG trajectory from that starting point runs into the broken 
(low-temperature) phase and for $\tau > 1$ the RG trajectory ends up in 
the symmetric (high-temperature) phase.
}
\label{fig3} 
\end{figure}

%===============
As a first step we apply the thermal RG equation \eq{thermal_opt_rg_k} to the $\phi^4$ 
(more precisely, to the $\phi^{2n}$) model in low dimensions. For the sake of simplicity 
we consider the following scalar potential,
\beq
V_k(\phi) = \sum_{n=1}^{\rm NCUT} \frac{g_{2n,k}}{(2n)!} \phi^{2n} 
\hskip 0.0cm \to \hskip 0.0cm
\tilde V_k(\tilde \phi) = \sum_{n=1}^{\rm NCUT} \frac{\tilde g_{2n,k}}{(2n)!} {\tilde \phi}^{2n} 
\eeq
where the RG scale-dependence is encoded in the couplings and we introduced 
dimensionless quantities denoted by the tilde superscript. The dimensionful RG flow 
equations for two couplings, i.e., for NCUT=2 have the following forms,
\begin{align}
\label{g2_flow}
k \partial_k & g_{2,k} = \frac{2 \alpha_{d-1}}{d-1} \, k^{d+1} 
\left[
	-\frac{g_{4,k} \coth \left(\frac{\sqrt{k^2+g_{2,k}}}{2 \tau k }\right)}
		{4 \left(k^2+g_{2,k}\right)^{3/2}} 
	\right. \nonumber \\ & \left. 
	-\frac{g_{4,k} \; \text{csch}^2\left(\frac{\sqrt{k^2+g_{2,k}}}{2 \tau k}\right)}
		{8 \tau k \left(k^2+g_{2,k}\right)} 
\right], \\
\label{g4_flow}
k \partial_k & g_{4,k} =  \frac{2 \alpha_{d-1}}{d-1} \, k^{d+1} 
\left[
	\frac{9 g_{4,k}^2 \coth \left(\frac{\sqrt{k^2+g_{2,k}}}{2 \tau k }\right)}
		{8 (k^2+g_{2,k})^{5/2}}
	\right. \nonumber \\ & \left. 
	+
	\frac{9 g_{4,k}^2 \text{csch}^2\left(\frac{\sqrt{k^2+g_{2,k}}}{2 \tau k}\right)}
		{16 \tau k (k^2+g_{2,k})^2}
	\right. \nonumber \\ & \left. 
	+
	\frac{3 g_{4,k}^2 \coth \left(\frac{\sqrt{k^2+g_{2,k}}}{2 \tau k }\right)
		\text{csch}^2\left(\frac{\sqrt{k^2+g_{2,k}}}{2 \tau k}\right)}
		{16 \tau^2 k^2 (k^2+g_{2,k})^{3/2}}
\right],
\end{align}
where $g_{4,k}$ and $g_{2,k}$ are dimensionful couplings. One can switch from 
dimensionful to dimensionless couplings by introducing $g_{2,k} = \tilde g_{2,k} k^2$ 
and $g_{4,k} = \tilde g_{4,k} k^{(4-d)}$. The RG flow equations given for dimensionless
couplings $\tilde g_{2,k}$ and $\tilde g_{4,k}$ have no explicit $k$-dependence. 
Thus, one can look for non-trivial, WF fixed point solutions. Indeed, \fig{fig2}
shows the thermal RG flow diagram of the $\phi^4$ model in $d=3$ dimensions.
%
% Fig 2
%
Empty circles show how the position of the WF fixed point changes with $\tau$.
The RG trajectory which runs from the Gaussian ($\tilde g_2 = 0$ and $\tilde g_4=0$) 
fixed point to the WF one separates the phases of the model. If one increases the 
value of $\tau$, the slope of the critical trajectory and consequently the broken phase 
is decreased. For $\tau \to \infty$ it is not possible to find any starting point in the 
vicinity of the Gaussian fixed point from which an RG trajectory can run into the 
broken phase. Since $\tau$ measures how thermal fluctuations are important 
compared to quantum fluctuations, one can say that by changing $\tau$ a thermal 
phase transition occurs. For an arbitrary but fixed starting point taken from the vicinity 
of the Gaussian fixed point, see the black cross on \fig{fig3} one can always 
determine a critical value $\tau_c$: if $\tau < \tau_c$ the RG trajectory from that 
starting point runs into the broken (low-temperature) phase and for $\tau > \tau_c$ 
the RG trajectory ends up in the symmetric (high-temperature) phase. 
%
% Fig 3
%

Let us consider the RG flow diagram shown in \fig{fig3} with respect to the change 
in the 'quantum' parameter which results in a QPT. The 'quantum' parameter is the
slope of RG trajectories taken at the vicinity of the Gaussian fixed point. For a fixed 
temperature, i.e., for fixed $\tau$, one can determine a critical value which is the 
slope of the separatrix, see dashed line of \fig{fig3}. If the slope of an RG trajectory 
is smaller or larger then this critical value, the corresponding trajectory runs into 
the broken or the symmetric phase respectively which signals the QPT.  
A good approximation for the critical value of the 'quantum' parameter is the ratio 
$(\tilde g_4/\vert \tilde g_2\vert)_{\rm{WF}}$ where $\tilde g_4$ and $\tilde g_2$ are 
the coordinates of the WF fixed point. Thus, the critical line on the QPT-CPT diagram 
of the $\phi^4$ model can be given in terms of $\tau_c$ and 
$(\tilde g_4/\vert \tilde g_2\vert)_{\rm{WF}}$ which are related to each other.

%===============
\section{QPT-CPT diagram of the $\phi^4$ model} 
\begin{figure}
\centering
\includegraphics[width=\linewidth]{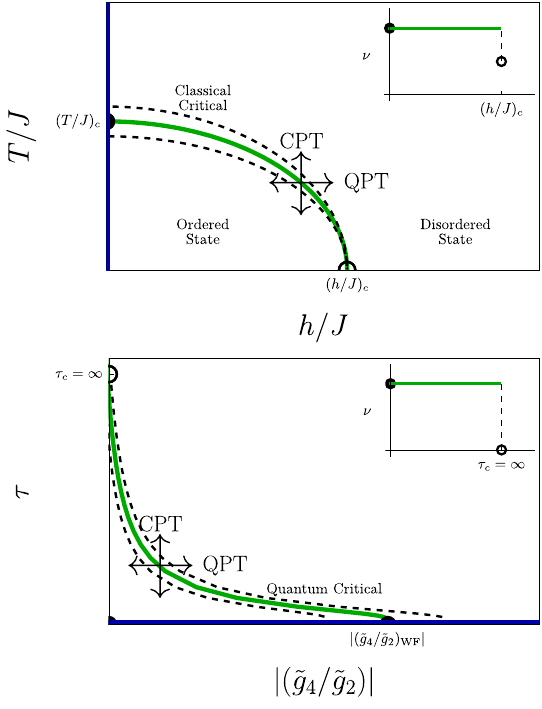}
\caption{
Schematic QPT-CPT diagram of the Ising model (to the top) and the $\phi^4$ 
theory (to the bottom) which are "dual to each other", i.e., $(T/J)$ corresponds 
to ($\tilde g_4/\tilde g_2$) while $\tau$ is related to $(h/J)$. Thus, the critical 
behaviour on blue lines known to be identical and expected to be the same
along the green critical lines but has a discontinuity at their endpoints. 
}
\label{fig4} 
\end{figure}

%===============
As a next step we discuss some key issues regarding the comparison of the 
QPT-CPT diagrams of the Ising model (which is well-known) and the $\phi^4$ 
theory (which we would like to map out here). The schematic pictures of these
QPT-CPT diagrams are shown in \fig{fig4}.
%
% Fig 4
%

It is known that the classical Ising model is equivalent to the zero-temperature $\phi^4$ 
QFT. Thus, the vertical axis of the he QPT-CPT diagram and the critical temperature of 
the Ising model can be mapped onto the horizontal axis of the he QPT-CPT diagram 
and the critical 'quantum' parameter of the (zero-temperature) $\phi^4$ QFT, see the
blue lines of \fig{fig4}. In addition, the thermal critical behaviour of the classical Ising 
model must be identical to the quantum critical behaviour of the $\phi^4$ QFT. 

It is clear from \fig{fig4}, that the 'quantum' parameter of the Ising model is $h/J$ which is the 
strength of the transverse-field in terms of $J$. The quantum critical behaviour of the $\phi^4$ 
QFT is given in terms of the irrelevant coupling around the WF saddle point, however, this 
irrelevant coupling is related to the ratio $(\tilde g_4/\vert \tilde g_2\vert)_{\rm{WF}}$ and in 
the previous section we argued, that this ratio can serve as a good 'quantum' parameter for 
$\phi^4$ QFT. Thus, if the temperature is zero, both the quantum Ising model and the 
$\phi^4$ QFT are non-trivial, they undergo a QPT with a finite $h_c$ and finite ratio 
$(\tilde g_4/\vert \tilde g_2\vert)_{\rm{WF}}$.

For vanishing quantum parameter, i.e., $h/J = 0$, the Ising model becomes the well-known 
classical one, which is non-trivial, it has a classical (thermal) phase transition with a finite 
transition temperature $T_c < \infty$. On contrary, for a vanishing 'quantum' parameter, i.e., 
for $(\tilde g_4/\vert \tilde g_2\vert)_{\rm{WF}} =0$, the $\phi^4$ model becomes a trivial 
massive free field theory (with thermal fluctuations), which has no phase transition, so there 
is no room for any finite transition temperature. On the one hand, the critical temperature of
the $\phi^4$ model must be zero if $(\tilde g_4/\vert \tilde g_2\vert)_{\rm{WF}} =0$. On 
the other hand, the critical line must be a strictly monotonic function, so one expects 
$T_c \to \infty$ for the $\phi^4$ in the limit $(\tilde g_4/\vert \tilde g_2\vert)_{\rm{WF}} \to 0$.
Therefore, one expects a discontinuity in the classical limit of the $\phi^4$ theory; this is
why we use an empty circle at $\tau_c = \infty$ on the lower panel of \fig{fig4} and in its
inset. In addition, one can conclude that the Ising model has a thermal (classical) and a 
quantum critical behaviour, too but the $\phi^4$ model has a quantum critical behaviour, only. 

Let us pay the attention of the reader that the terminology 'classical' has to be used with great 
care. In this work when we use 'classical Ising model' we refer to the Ising model with thermal 
fluctuations, however, when we use 'classical' $\phi^4$ theory it is understood as a model with no 
quantum and thermal fluctuations. The change in the sign of the mass parameter of the classical 
$\phi^4$ theory results in two different (single and double well) form for the potential but actually 
it is not considered as a phase transition. As we argued, the 'quantum' parameter of the $\phi^4$ 
QFT is the irrelevant direction around the WF fixed point which is a combination of the quartic 
and quadratic couplings. Thus, if one takes the limit of vanishing 'quantum' parameter (in some 
sense the classical limit) one arrives at a free field theory (with no quartic term) which is not 
equivalent to the 'classical' $\phi^4$ model.

It is also known that the critical exponents of the finite temperature quantum Ising model 
are equal to the critical exponents of the classical Ising model, see the inset of the upper
panel of \fig{fig4}. Since the Ising model and the $\phi^4$ theory are "dual to each other", 
i.e., $(T/J)$ corresponds to ($\tilde g_4/\tilde g_2$) while $\tau$ is related to $(h/J)$, the 
critical behaviour is expected to be the same along the green critical lines of \fig{fig4} and
the critical exponents of the finite temperature $\phi^4$ QFT must be equal to their 
zero-temperature values, see the inset on the lower panel of \fig{fig4}. Thus, for finite 
temperatures and finite 'quantum' parameters, the Ising and the $\phi^4$ models are 
expected to belong to the same universality class. 

However, it is well-known that the critical behaviour of the Ising model has a discontinuity 
in the limit $T \to 0$, see the empty circle at $T=0$ (i.e., at $h_c$) on the upper panel of 
\fig{fig4}. So, critical lines (green lines on the upper and lower panels of \fig{fig4}) have a 
discontinuity at their endpoints: for the Ising it happens at $h/J = (h/J)_c$ and for the 
$\phi^4$ theory it situates at $\tau = \tau_c =\infty$.

These conclusions can be drawn without performing any explicit calculations on the 
QPT-CPT diagram of the $\phi^4$ model. In other words, one can assume the above 
statements which can be used to test the thermal RG equations \eq{thermal_opt_rg_k} 
and \eq{thermal_sharp_rg_k} suggested by us. 

\begin{figure}
\centering
\includegraphics[width=\linewidth]{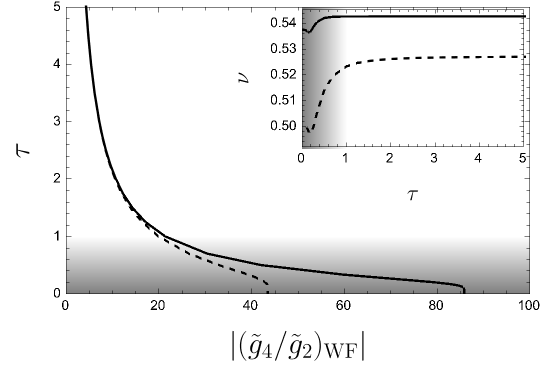}
\caption{
QPT-CPT diagram of the $\phi^4$ model in $d=3$ dimensions based on the thermal RG 
equations \eq{thermal_opt_rg_k} and \eq{thermal_sharp_rg_k}. The solid (dashed) line 
stands for results obtained by the Litim (sharp-cutoff) regulator. The inset shows the 
critical exponent $\nu$ as a function of $\tau$ which is related to the temperature. The 
shaded areas stand for $\tau<1$.}
\label{fig5} 
\end{figure}

The QPT-CPT diagram of the three-dimensional $\phi^4$ model is constructed by the numerical 
solution of the thermal RG equations \eq{thermal_opt_rg_k} and \eq{thermal_sharp_rg_k},
see the solid (Litim cutoff) and the dashed (sharp-cutoff) lines of \fig{fig5}.
%
% Fig 5
%
The QPT-CPT diagram based on thermal RG equations with the Litim and the sharp-cutoff 
regulators are similar but not identical: between $\tau =1$ and $\tau= \infty$ they 
coincide but differ from each other between $\tau =0$ and $\tau= 1$. 

The critical exponent $\nu$ of the $\phi^4$ model is expected to show a small dependence 
on the parameter $\tau$, practically they must be constant along the critical lines which are 
see the solid (Litim regulator) and dashed (sharp cutoff regulator) lines of \fig{fig5}. Indeed, 
the solid and the dashed lines of the inset on \fig{fig5} confirms this. Between $\tau=1$ and 
$\tau =\infty$ the solid line is practically constant and it has a relatively small dependence 
between $\tau=0$ and $\tau =1$. The dashed line on the inset of \fig{fig5} shows that by 
using the thermal RG equation with the sharp-cutoff regulator, the critical exponent $\nu$ 
has a larger dependence on the parameter $\tau$, indeed $\nu$ varies with $\tau$ even in 
the range $\tau>1$ which was not true for the Litim case (solid line). Nevertheless, one 
can argue that the Limit regulator always provides us a better convergence and as a 
consequence, more reliable results in the zero-temperature limit, i.e., for $\tau \to 0$.
However, this requires further investigations by going beyond LPA and by the study of 
different models, such as the two-dimensional sine-Gordon theory.
In both cases, i.e., for the Litim and the sharp-cutoff regulators the critical exponent $\nu$ 
tends to a constant value in the limit $\tau \to \infty$ which is identical to that of obtained by 
the zero-temperature FRG equation. Of course, this constant value depends on the choice 
of the regulator. Let us draw the attention of the reader that the Litim regulator is known to 
give the closest result among all regulators to the "exact" value ($\nu = 0.629971$).

It is also expected that the critical line of the $\phi^4$ model must tend to infinity in the 
limit of vanishing quantum parameter which is clearly supported by the thermal RG studies, 
see the solid and dashed lines of \fig{fig5}. Our results are tested for larger values 
of NCUT and we have found the same qualitative behaviour. 

\begin{figure}
\centering
\includegraphics[width=\linewidth]{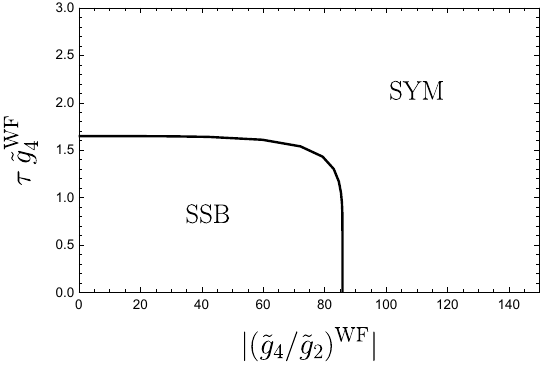}
\caption{
QPT-CPT diagram of the $\phi^4$ model in $d=3$ dimensions in terms of $\tau \tilde g_4^{\rm WF}$ 
and $(\tilde g_4/\tilde g_2)^{\rm WF}$ where $\tilde g_4^{\rm WF}$ and $\tilde g_2^{\rm WF}$ are the coordinates 
of the WF fixed point. The black critical line which separates the spontaneous symmetry broken (SSB) 
and the symmetric (SYM) phases terminates at finite value in the limit of $(\tilde g_4/\tilde g_2)^{\rm WF} \to 0$ 
which represents an important difference compared to \fig{fig5} where the critical line runs to infinity. }
\label{fig6} 
\end{figure}

\begin{figure}
\centering
\includegraphics[width=\linewidth]{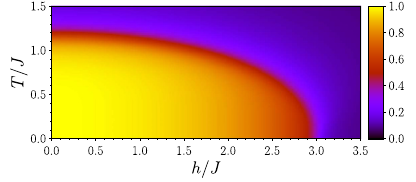}
\caption{
QPT-CPT diagram of the transverse field Ising model obtained by Monte Carlo simulation
and taken from \cite{qpt_cpt_transverse_field_ising}. This can be directly compared to 
\fig{fig6}.}
\label{fig7} 
\end{figure}

Finally let us discuss an interesting property of the infinity temperature limit. 
We showed, that in the limit $\tau \to \infty$ the WF fixed point tends to the horizontal axis.
In other words, the quartic coupling $\tilde g_4^{\rm WF}$ of the WF fixed point tends to zero 
for $\tau \to \infty$.  However, one can show that their product, $\tau \tilde g_4^{\rm WF}$ 
tends to a constant value. Thus, the QPT-CPT diagram where the vertical axis is chosen
to be $\tau \tilde g_4^{\rm WF}$, see \fig{fig6}, becomes identical to the well-known QPT-CPT 
diagram of the Ising model obtained by Monte Carlo simulation, see \fig{fig7}.
%
% Fig 6
%
It is important to note that real (and not pseudo) fixed points can only
be determined by keeping $\tau = T/k$ fixed over the RG flow in the simultaneous 
$T, k \to 0$ limit. This provides us the tool to draw the QPT-CPT diagram of a QFT,
such as the $\phi^4$ model, see \fig{fig6}. The critical line which separates the broken
and the symmetric phases is determined by these (real) fixed points
which can be directly compared to simulation results and measurements performed on
the corresponding spin model, in our case on the transverse field Ising model, see \fig{fig7}. 
If one keeps $T$ constant over the RG flow no such comparison can be done since in this case 
the thermal FRG has pseudo (and not real) fixed points. So, the direct comparison between 
the line of real fixed points of quantum-thermal spin models and their thermal QFT counterparts 
can only be done in the modified thermal FRG method proposed here.
%
% Fig 7
%
%

%===============
\section{Conclusions}
%===============
In this work, we proposed a modified finite-temperature FRG approach by relating the 
temperature parameter to the running RG scale, $T \equiv k_T = \tau k$ where $k_T$ 
is the running cutoff for thermal, and $k$ is for the quantum fluctuations. In this case, 
the temperature must be related to the dimensionless quantity $\tau$. This choice is 
supported by the idea of the Wilsonian approach where fluctuations are taken into 
account by the integration of degrees of freedom above these running cutoffs which 
are chosen to be different in order to make difference between thermal and quantum 
fluctuations. For $k_T = k$ (i.e.,~for $\tau =1$) one cannot distinguish between thermal 
and quantum fluctuations. We applied this new thermal RG approach for the $\phi^4$ 
model in lower dimensions and constructed its QPT-CPT phase diagram which was 
used to test the new thermal RG approach. Indeed, we formulated requirements for 
the QPT-CPT phase diagram of the $\phi^4$ theory based on known properties of the 
QPT-CPT phase diagram of the Ising model and we used these requirements to prove 
the viability of the thermal RG method proposed in this work.

It is important to note that previous studies based on the finite-temperature FRG method
discussed findings with special emphasis on the interplay of the temperature $T$, 
and the running RG scale $k$. Above Eq.~(58) of \cite{thermal_rg_phi4_volume} one finds 
for example, the RG flow equation of the infinite volume Stefan--Boltzmann pressure 
$\partial_k P(T) = -k^4/12\pi^2 [\coth(k/2T) -1]$. To draw conclusions on, one has to discuss 
how the running RG scale $k$ is related to the temperature which is fixed to the UV cutoff,
i.e., $T = \tau \Lambda$. For large cutoff scales, $k >> T$ the flow decays exponentially 
with $\exp(-k/T)$ which is given by Eq.~(58) of \cite{thermal_rg_phi4_volume}. By using 
our suggestions where the temperature parameter is related to the running scale $T = \tau k$,
and the dimensionless parameter $\tau$ plays the role of the temperature, this flow equation 
is simplified as $\partial_k P(\tau) = -k^4/12\pi^2 [\coth(1/2\tau) -1]$ and conclusions can be drawn
based on the change of the dimensionless parameter $\tau$ only.
Thus, the modified thermal
RG method presented in this work finds application in a natural way in all previous thermal RG 
studies of quantum models and it can be used both to simplify the treatment and find new results 
in a variety of quantum models exhibiting quantum phase transitions.

%===============
\section*{Acknowledgements} 
%===============
Support by the UD Program for Scientific Publication and the CNR/MTA 
Italy-Hungary 2023-2025 Project "Effects of strong correlations in interacting 
many-body systems and quantum circuits" is acknowledged. 
The creation of this scientific communication was supported by the University of Miskolc 
with funding granted to the author I.~N\'andori within the framework of the 
institution's Scientific Excellence Support Program. (Project identifier: ME-TKTP-2025-3).
We thank Eszter Napsug\'ar Nyergesy for discussions and for the careful reading 
of the MS.


\begin{thebibliography}{97}

\bibitem{sachdev}
S. Sachdev, {\it Quantum phase transitions} (Cambridge, Cambridge University Press, 2011).
 
\bibitem{tranverse_ising_1}
R. J. Elliott and C. Wood, J. Phys. C: Solid State Phys. {\bf 4} 2359 (1971).

\bibitem{tranverse_ising_2}
P. Pfeuty and R. J. Elliott, J. Phys. C: Solid State Phys. {\bf 4}, 2370 (1971);
P. Pfeuty, Ann. Phys. {\bf 57}, 79 (1970).

\bibitem{shankar}
R. Shankar, {\it Quantum field theory and condensed matter: an introduction} (Cambridge, Cambridge University Press, 2017).
  
\bibitem{eea_rg}
C. Wetterich, Phys. Lett. B {\bf 301}, 90 (1993);  
T. R. Morris, Int. J. Mod. Phys. A {\bf 9}, 2411 (1994).

%%%%%%%%%%%%%%%%%%%

\bibitem{rg_dimful_T_1}
N. Tetradis and C. Wetterich, Nucl. Phys. B {\bf 398}, 659 (1993).

\bibitem{rg_dimful_T_2}
M. Pietroni, N. Rius, and N. Tetradis, Phys. Lett. B {\bf 397}, 119 (1997).

\bibitem{finite_T_V_1}
J. Braun, B. Klein, and H. Pirner, Phys. Rev. D {\bf 72}, 034017 (2005).

\bibitem{thermal_rg_phi4}
J.-P. Blaizot, A. Ipp, R. M\'endez-Galain, and N. Wschebor, Nucl. Phys. A {\bf 784}, 376 (2007). 

\bibitem{thermal_rg_phi4_pressure}
J.-P. Blaizot, A. Ipp, and N. Wschebor, Nucl. Phys. A {\bf 849} 165 (2011).

\bibitem{rg_dimful_T_5}
L. Fister and J. M. Pawlowski, \verb|arXiv:1112.5440|% (2011).

\bibitem{rg_dimful_T_6}
L. Fister and J. M. Pawlowski, PoS QCD-TNT-II, 021 (2011).

\bibitem{rg_dimful_T_8}
J. Braun, S. Diehl, and M. M. Scherer, Phys. Rev. A {\bf 84}, 063616 (2011).

\bibitem{finite_T_V_3}
J. Braun, B. Klein, and B.-J. Schaefer, Phys. Lett. B {\bf 713}, 216 (2012).

\bibitem{rg_dimful_T_10}
L. Fister and J. M. Pawlowski, Phys. Rev. D {\bf 88}, 045010 (2013).

\bibitem{finite_T_V_5}
R.-A. Tripolt, J. Braun, B. Klein, and B.-J. Schaefer, Phys. Rev. D {\bf 90}, 054012 (2014).

\bibitem{thermal_rg_phi4_volume}
L. Fister and J. M. Pawlowski, Phys. Rev. D {\bf 92}, 076009 (2015).

%%%%%%%%%%%%%%%%%%%%%%%%

\bibitem{rg_dimless_T_1}
J. Braun and H. Gies, JHEP {\bf 0606}, 024 (2006).

\bibitem{rg_dimless_T_2}
P. Jakubczyk, P. Strack, A.A. Katanin, and W. Metzner, Phys. Rev. B {\bf 77}, 195120 (2008).

\bibitem{rg_dimless_T_3}
J. Braun, Eur. Phys. J. C {\bf 64} 459 (2009). 

\bibitem{rg_dimless_T_4}
P. Jakubczyk, Phys. Rev. B {\bf 79}, 125115 (2009).

\bibitem{rg_dimless_T_5}
P. Strack, and P. Jakubczyk, Phys. Rev. B {\bf 80}, 085108 (2009).

\bibitem{rg_dimless_T_6}
P. Jakubczyk, W. Metzner, and H. Yamase, Phys. Rev. Lett. {\bf 103}, 220602 (2009).

\bibitem{rg_dimless_T_7}
P. Jakubczyk, J. Bauer, and W. Metzner, Phys. Rev. B {\bf 82}, 045103 (2010).

\bibitem{rg_dimless_T_8}
J. Braun, B. Klein, and P. Piasecki, Eur. Phys. J. C {\bf 71}, 1576 (2011).

\bibitem{rg_dimless_T_9}
J. Braun. J. Phys. G {\bf 39}, 033001 (2012).

\bibitem{rg_dimless_T_10}
D. D. Scherer, J. Braun, and H. Gies, J. Phys. A: Math. Theor. {\bf 46} 285002 (2013).

\bibitem{rg_dimless_T_11}
J. Braun, M. Leonhardt, and M. Pospiech, Phys. Rev. D {\bf 96}, 076003 (2017).

%\bibitem{scherer_24} B. Hawashin, J. Rong, and M. M. Scherer, \verb|arXiv:arXiv:2409.10606|

%%%%%%%%%%%%%%%%%%%

\bibitem{rancon_1}
A. Rancon and N. Dupuis, Phys. Rev. B {\bf 84}, 174513 (2011).  

\bibitem{rancon_2}
A. Rancon and N. Dupuis, , Phys. Rev. B {\bf 83}, 172501 (2011).

\bibitem{rancon_3}  
 A. Rancon, O. Kodio, N. Dupuis, and P. Lecheminant, Phys. Rev. E {\bf 88}, 012113 (2013).

%%%%%%%%%%%%%%%%%%%

\bibitem{BrBrCoDa}
C. Branchina, V. Branchina, F. Contino, and N. Darvishi, Phys. Rev. D {\bf 106}, 065007 (2022).

\bibitem{BrBrCo}
C. Branchina, V. Branchina, and F. Contino, Phys. Rev. D {\bf 107}, 096012 (2023).

%%%%%%%%%%

\bibitem{qpt_higgs1}
T. Steingasser, \verb|arXiv:2405.02415|%, DOI: https://doi.org/10.22323/1.463.0150

\bibitem{qpt_higgs2}
T. Steingasser and I. Kaiser, Phys. Rev. D {\bf 108}, 095035 (2023).

%%%%%%%%%%

\bibitem{wilson}
K. G. Wilson, Phys. Rev. B {\bf 4}, 3174 (1971); {\it ibid.} 
%K. G. Wilson, Physical Review B
{\bf 4}, 3184 (1971).

%%%%%%%%%%%%%%%%%%%

\bibitem{Litim2000}
D. F. Litim, Phys. Lett. B {\bf 486}, 92 (2000).

\bibitem{wh_rg}
F. J. Wegner and  A. Houghton, Phys. Rev. A. {\bf 8}, 401 (1973). 

\bibitem{thermal_rg_cosmo} 
I. G. M\'ari\'an, A. Trombettoni, and I. N\'andori, Phys. Lett. B {\bf 858}, 139051 (2024).
 
\bibitem{cylindrical_symm_1}
I. Steib, S. Nagy, and J. Polonyi, Int. J. Mod. Phys. A {\bf 36}, 2150031 (2021).
  
\bibitem{cylindrical_symm_2}  
F. G\'eg\'eny and S. Nagy, Int. J. Mod. Phys. A {\bf 36}, 2250061 (2022).

%%%%%%%%%%%%%%%%%%%

\bibitem{negele}
J. W. Negele and H. Orland, {\it Quantum many-particle systems} (Reading, Perseus, 1998).

\bibitem{kapusta}
J. I. Kapusta, {\it Finite-temperature field theory: principles and applications} (Cambridge, Cambridge University Press, 2023).

\bibitem{qpt_cpt_transverse_field_ising}
B. Blass, H. Rieger, {\it Test of quantum thermalization in the two-dimensional transverse-field Ising model}, 
Sci. Rep. {\bf 6} 38185, (2016).

\bibitem{details}
For example, in \cite{rg_dimless_T_9}
at page 42, in Eq.(132) one finds $\partial_t \tau = - \tau$ where $t = \ln(k/\Lambda)$ and
in \cite{rg_dimless_T_11} at the top of the page 12 one can read $\partial_t T/k = - T/k$ 
which result in the trivial RG scale dependence for $\tau_k$. Another example is found in 
\cite{rg_dimless_T_7} at page 4 above equation (21), $\tilde T(\Lambda^\star) = 1$ where 
the intermediate momentum scale $\Lambda^\star$ is fixed by the running dimensionless 
temperature $\tilde T$ which is identical to $\tau$. Finally, we refer to \cite{rancon_3} at 
page 7 right below equation (50), where one finds $\tilde T_k = T/(c_k k)$. 

\end{thebibliography}
\end{document}